\documentclass[showpacs,showkeys,preprintnumbers,pre,amsmath,amssymb,superscriptaddress,twocolumn]{revtex4-2}
\usepackage[english]{babel}
\usepackage[utf8]{inputenc}
\usepackage[colorinlistoftodos, color=green!40, prependcaption]{todonotes}

\usepackage[colorlinks=true,linkcolor=blue,urlcolor=blue,citecolor=blue]{hyperref}
\usepackage[normalem]{ulem}

\DeclareUnicodeCharacter{2212}{-}

\begin{document}
\title{Stochastic Quantization of General Relativity \`a la Ricci-Flow}

\author{Matteo Lulli}
\email{mlulli@phy.cuhk.edu.hk}
\address{Department of Physics, The Chinese University of Hong Kong, Sha Tin, Hong Kong, China}

\author{Antonino Marcian\`o}
\email{marciano@fudan.edu.cn}
\affiliation{Center for Field Theory and Particle Physics \& Department of Physics, Fudan University, 200433 Shanghai, China}
\affiliation{Laboratori Nazionali di Frascati INFN, Frascati (Rome), Italy, EU}
\affiliation{INFN sezione Roma Tor Vergata, I-00133 Rome, Italy, EU}

\author{Xiaowen Shan}
\email{xiaowenshan@uic.edu.cn}
\affiliation{Institute of Advanced Study, BNU-HKBU United International College, Zhuhai, Guangdong 519088, China}

\begin{abstract}
\noindent 
We follow a new pathway to the definition of the Stochastic Quantization (SQ), first proposed by Parisi and Wu, of the action functional yielding the Einstein equations. Hinging on the functional similarities between the Ricci-Flow equation and the SQ Langevin equations proposed by Rumpf, we push forward a novel approach characterized by a multiplicative noise and a stochastic time that converges to the proper time of a space-like foliation in the equilibrium limit, where quantities have constant averages. We express the starting system of equations using the Arnowitt-Deser-Misner (ADM) variables and their conjugated Hamiltonian momenta. Such a choice is instrumental in understanding the newly derived equations in terms of the breakdown of the diffeomorphism invariance of the classical theory, which instead will hold on average at the steady state. We comment on the physical interpretation of the Ricci flow equations, and argue how they can naturally provide, in a geometrical way, the renormalization group equation for gravity theories. In the general setting, the equation associated to the shift vector yields the Navier-Stokes equation with a stochastic source.
Moreover, we show that the fluctuations of the metric tensor components around the equilibrium configurations, far away from the horizon of a Schwarzschild black hole, are forced by the Ricci flow to follow the Kardar-Parisi-Zhang equation, whose probabilistic distribution can yield an intermittent statistics. We finally comment on the possible applications of this novel scenario to the cosmological constant, arguing that the Ricci flow may provide a solution to the Hubble tension, as a macroscopic effect of scale dependence of the quantum fluctuations of the metric tensor. 
\end{abstract}
\maketitle

\section{Introduction}
\noindent 
Quantizing gravity has become a longstanding problem, posing continuous challenges over the last 100 years, i.e. since the birth of General Relativity (GR). Many similarities exist between GR and other gauge theories that can be recovered casting the Einstein-Hilbert action in terms of the field strength of the gravitational spin-connection and the tetrad as frame-field. On one side, the Hamiltonian first-class constraints of GR provide its classical equations of motion, together with the other Hamilton equations, generating space-time diffeomorphisms; on the other side, for non-Abelian Yang-Mills theories, the (Gauss) constraint emerges as a subset of the classical equations of motion, providing the only constraints to the theory and generating gauge transformations. While these latter are not observable, space-time reparametrizations act in a way that can be measured experimentally. 
From a different perspective, the Hamiltonian path integral approach to GR enforces the constraints in the definition of the measure by means of functional deltas \cite{henneaux}, similarly to the Faddeev-Popov gauge fixing for Yang-Mills gauge theories.
    
Because of the different nature of the symmetries implemented through the constraints, it immediately appears that while for Yang-Mills gauge theories the quantization procedure allows for some \emph{quantum oscillations} around the saddle point of the classical equations of motion, the same does not hold true for GR. 
The imposition of the constraints in the integration measure does not allow the integral to \emph{sample} possible fields configurations that do not belong to the saddle point. In other words, configurations that do not obey the classical equations of motion, i.e. four out of ten Einstein equations corresponding to first-class constraints, are forbidden. This is in stark contrast with the commonly accepted interpretation, for which the path integral is exactly a way to weight the quantum contributions (of the fields configuration) that do not belong to the saddle point. Thus, from the perspective of the symmetries, the usual path integral quantization of GR always considers only classical fields configurations. \\

From the Lagrangian perspective, the path-integral formulation of gauge theories is not implemented as a sum over quantum paths fluctuating away from the constraints' hyper-surfaces. Instead, gauge fixing is encoded in the path integral in a way that renders manifestly invariant the quantization of the theory from the particular gauge-fixing condition that is chosen. But a similar procedure cannot be applied to gravity. Constraints of this latter generate \emph{external symmetries}, for which the gauge-fixing procedure, and hence the invariance of the path-integral quantization on the gauge-fixing, becomes meaningless.\\

From the Hamiltonian perspective, the canonical path integral is typically formulated by introducing, in the functional measure, generalized deltas that strictly enforce first-class constraints which would otherwise appear as equations of motion at the saddle point. Hence, from the canonical perspective it is not possible to sample fields configurations along trajectories that do not break general covariance. However, the constraints, being equations of motion themselves, only generate the symmetries for the trajectories at the saddle point and not away from it. Hence, this procedure greatly reduces the space of available fields configurations for the quantization and does not allow for discussing a possible emergence of general covariance in the classical limit.\\

It is appealing to look for different approaches that assume, instead, an explicit violation of the whole set of equations of motion, i.e. for a path integral sampling entirely away from the saddle point. It turns out that this is the main feature of the Stochastic Quantization method~\cite{Parisi_Wu_1981}, in which quantization is performed by reaching the steady-state of a stochastic process modelled by a Langevin equation for the fields. This latter is customarily expressed, over a manifold with Euclidean signature, as  
\begin{equation}
\frac{\partial}{\partial s}\phi_{A}\left(x^{\mu},s\right)=-\frac{\delta S\left[\phi\right]}{\delta\phi_{A}}+\eta_A\left(x^{\mu}, s\right)\,,
\end{equation}
where the drift term is provided by the first variation of the action functional $S$, and $\eta_A$ is the associated additive noise with $\langle\eta_{A}\left(x,s\right)\eta_{B}\left(x',s'\right)\rangle=G_{AB}\delta\left(x-x'\right)\delta\left(s-s'\right)$, where $G_{AB}$ is some covariance matrix for the noise of the different fields. The evolution is computed with respect to a fictive stochastic time $s$, which is generally treated as an extra (non-physical) dimension with respect to the chosen ambient space, e.g. the Minkowskian four-dimensional space-time. The stochastic equations, either interpreted in the It\^o or Stratonovich way, allow one to write the related evolution equation for the probability distribution of the fields, i.e. the Fokker-Planck equation~\cite{gardiner2004handbook}.\\

At this point, it is possible to show that the stationary distribution is largest for field configurations yielding diffeomorphism symmetry. Instantaneous configurations will likely break the diffeomorphisms algebra, but the latter is expected to hold on average in the equilibrium long-stochastic-time limit. 
The breaking of symmetries at a given short (stochastic) time scale, which are then recovered on larger time scales, is something very similar to what happens in turbulence~\cite{Frisch_1995}, for which the Galilean symmetries, broken at small scales by the intermittent fluctuations of the velocity field, are recovered on large (spatial) scales.\\ 

In this paper we propose a new take on the SQ programme applied to GR. We propose two main modifications: (i) the exchange of the usual additive noise term with a \emph{multiplicative} one; (ii) a geometric definition of the stochastic time which is promoted from a mere extradimensional
fictitious parameter to a physical quantity that is proportional to the proper time at equilibrium, thus behaving as a scalar under diffeomorphisms. These two assumptions lead to a number of intertwined consequences.\\ 

As shown below, close to equilibrium, where averages are constant in the thermal time, the stochastic time $s$ we introduce is directly related to a rescaling of the metric tensor as a function of the Jacobian of the transformation between $s$ and the proper time $\tau$. The out-of-equilibrium relaxation of the stochastic process can be consistently related to a space-time dependent scale transformation, which can be interpreted as generating the Renormalization Group flow of an effective action. Such effective action is defined by the saddle point of the probability distribution solution of the finite (stochastic) time Fokker-Planck equation associated to the multiplicative random process. On the other hand the multiplicative noise naturally induces the emergence of a cosmological constant term related to the well-known shift (proportional to the square of the noise amplitude) of the saddle point with respect to the additive noise case. This result allows to interpret the cosmological constant as a macroscopic manifestation of quantum fluctuations of the gravitational field.\\

We will demonstrate the fundamental results we spelled in this introduction by (i) rewriting the system of equations in terms of the ADM variables and conjugate momenta and by applying them to (ii) the spherically symmetric and stationary space-time and (iii) to a metric describing the evolution of a isotropic and homogeneous universe. The use of the ADM variables (i) allows to highlight the breaking of the diffeomorphism algebra in the out-of-equilibrium regime while introducing a non-trivial eigenvalue of the super-Hamiltonian, thus providing a natural solution to the issue of the frozen formalism~\cite{CQG_2014} in the equilibrium limit. It also allows to emphasize a clear connection between the out-of-equilibrium relaxation of the super-momentum constraint and the Navier-Stokes equations with random forcing. Further, when considering the \emph{black-hole}-like metric we will show that, by virtue of the multiplicative noise and the implementation of Rumpf's condition for the compatibility with the It\^o calculus~\cite{Rumpf_1986}, the out-of-equilibrium dynamics of the \emph{lapse} function is described in terms of the Kardar-Parisi-Zhang equation. Especially, this finding not only highlights a rigorous mathematical connection with the dynamics of interface growth and that of Burgers' equation, but also explicitly introduces the concept of \emph{intermittency} of the quantum fluctuations of the metric tensor, thus providing a distinctive element of the gravitational field with respect to other field theories. Finally, the analysis of the cosmological case allows us to draw an enticing link between the out-of-equilibrium dynamics and the evolution of the effective value of the cosmological constant, which can then be leveraged in order to provide a fresh perspective on the problem of the Hubble tension \cite{DiValentino_2021}.\\

The plan of the paper is as follows. 
In Sec.~\ref{rfsq} we start by reminding the main features of the stochastic quantization procedure, and connects it to the the Ricci-flow equations. In Sec.~\ref{SSFP} we investigate the stationary regime of the Ricci flow equation and derive the related Fokker-Planck probability distribution. We then comment on the features of the saddle point configurations, which allow to recover the classical equations of motion of gravity, emphasizing the appearance of an effective cosmological constant that is connected to the amplitude of the stochastic noise of the Langevin equation. Finally, we ponder the choice of the stochastic time and shed light on the link with the conformal symmetry. In Sec.~\ref{results} we provide the main results we obtained in re-writing the Ricci equation in the Hamiltonian formalism. 
In Sec.~\ref{PI} we comment on the physical interpretation of the results we obtained, and in Sec.~\ref{CC} we apply the framework we have been developing to the study case of the running of the cosmological constant. Finally in Sec.~\ref{CO} we provide some preliminary conclusions and outlooks.

\section{From Ricci-flow to Stochastic Quantization} \label{rfsq}
\noindent
We start our analysis by deriving a generalized Langevin equation for the gravitational field that is related to the Ricci-flow.

The Ricci-flow equations~\cite{Hamilton_1982, Hamilton_1986, Hamilton_1993, Hamilton_1995, Perelman_entropy} have been historically cast for three-dimensional Riemannian manifolds as
\begin{equation}
\frac{\partial}{\partial s}g_{\mu\nu}=-2R_{\mu\nu}\,.
\label{eq:rf}
\end{equation}
A fundamental aspect of these equations is that, when considering the four-dimensional pseudo-Riemannian case, the fixed point, where $\partial g_{\mu\nu}/\partial s=0$, corresponds to the  Einstein vacuum equations. In particular one can easily see that the supplemental thermal time variable $s$ labels a sequence of manifolds that are not related by diffeomorphisms. Hence, general covariance is broken in the relaxation regime, while it is recovered at the fixed point.\\

The Ricci-flow equations drive the metric tensor 
according to a ``diffusive'' dynamics, and the flow they determine has been shown to be a gradient flow. The Langevin equation generalizes the concept of gradient flow by introducing a suitable noise term such that a potential function is minimized on average (when considering additive noise). This very principle is used as a building block for the Stochastic Quantization procedure proposed by Parisi and Wu \cite{Parisi_Wu_1981} and extended in~\cite{Rumpf_1986} to the gravitational field described, over manifolds with Lorentzian signature, by the equations
\begin{equation}
\frac{\partial g_{\mu\nu}\left(x,s\right)}{\partial s}=\imath\mathcal{G}_{\alpha\beta\mu\nu}\frac{\delta S}{\delta g_{\alpha\beta}}+\eta_{\mu\nu}\left(x,s\right)\,,
    \label{eq:sq_rumpf}
\end{equation}
where the super-metric $\mathcal{G}$ is defined as
\begin{equation}
\begin{split} & \mathcal{G}_{\alpha\beta\mu\nu}\left(x,x';\lambda\right)\\
= & \frac{2\kappa}{\sqrt{|g|}}\!\left[g_{\alpha\mu}g_{\beta\nu}+g_{\alpha\nu}g_{\beta\mu}-\frac{\lambda}{2\lambda+1}g_{\alpha\beta}g_{\mu\nu}\right]\!\delta^{\left(4\right)}\left(x-x'\right),
\end{split}
\end{equation}
with $\lambda \neq -1/2$, in four space-time dimensions~\cite{hawking1980general}, and $\kappa = 8\pi G/c^3$, where $G$ and $c$ are the Newton's constant and the speed of light, respectively. The parameter $\lambda$ plays a crucial role not only in the seminal work on the SQ of General Relativity~\cite{Rumpf_1986} but also in more recent non-perturbative approaches to quantum gravity~\cite{Horava_2009, Horava_2009_1}. Finally the noise variance was defined in Ref.~\cite{Rumpf_1986} as
\begin{equation}
\langle\eta_{\mu\nu}\left(x,s\right)\eta_{\rho\sigma}\left(x',s'\right)\rangle=\frac{2}{\kappa^2}\langle\mathcal{G}_{\mu\nu\rho\sigma}\left(x,x'\right)\rangle\delta\left(s-s'\right).
\label{eq:RumpfVariance}
\end{equation}

Let us now continue by drawing a closer analogy between Eq.~\eqref{eq:rf} and Eq.~\eqref{eq:sq_rumpf}. It is possible to rewrite the Ricci-flow equations as
\begin{equation}\label{fromRicci}
\begin{split}\frac{\partial}{\partial s}g_{\mu\nu} & =-2R_{\mu\nu}\\
 & =-2\left[R_{\mu\nu}-\frac{1}{2}g_{\mu\nu}R\right]-g_{\mu\nu}R \,.\\
\end{split}
\end{equation}
When comparing Eq.~\eqref{eq:rf} and Eq.~\eqref{eq:sq_rumpf}, a part from the imaginary unit $\imath$ and the specific choice $\lambda=0$, one is tempted to draw an analogy between the noise term $\eta_{\mu\nu}$ and $g_{\mu\nu} R$. The latter term suggests to consider a \emph{multiplicative} noise, i.e. $g_{\mu\nu} \,\eta(s)$ where $\eta$ has the same dimensions as $R$ as well as the same transformation properties under diffeomorphism. \\

Based on the aforementioned analogy, we propose the following {\it Ansatz}
\begin{equation}
\frac{\partial g_{\mu\nu}\left(x,s\right)}{\partial s}=\imath\mathcal{G}_{\alpha\beta\mu\nu}\frac{\delta S}{\delta g_{\alpha\beta}}+g_{\mu\nu}\left(x,s\right)e^{\imath\frac{\gamma}{2}}\sqrt{2\Lambda}\tilde{\eta}\left(x,s\right),
\label{eq:SQ_mult}
\end{equation}
where we identified the complex noise with
\begin{equation}
    \eta\left(s\right)=e^{\imath\frac{\gamma}{2}}\sqrt{2\Lambda}\, \tilde{\eta}\left(s\right),
\end{equation}
where both $\Lambda$ and the noise $\tilde{\eta}$ are real, so that the noise $\eta$ is made to be complex only by means of the arbitrary constant phase $\gamma$. The correlator of $\tilde{\eta}$ amounts to
\begin{equation}
    \langle\tilde{\eta}\left(s\right)\tilde{\eta}\left(s'\right)\rangle = \delta\left(s-s'\right),
\end{equation}
i.e., one obtains the differential of the Wiener process as $\tilde{\eta}\left(s\right)\text{d}s=\text{d}W\left(s\right)$.
It is interesting to notice that the framework proposed in~\cite{Rumpf_1986} can be specialized to the present case as 
\begin{equation}
\begin{split} & 2\Lambda e^{\imath\gamma}\langle g_{\alpha\beta}\left(s\right)g_{\mu\nu}\left(s'\right)\tilde{\eta}\left(s\right)\tilde{\eta}\left(s'\right)\rangle=\\
 & 2\Lambda e^{\imath\gamma}\,\mathcal{M}_{\alpha\beta\mu\nu}\langle\tilde{\eta}\left(s\right)\tilde{\eta}\left(s'\right)\rangle=2\Lambda e^{\imath\gamma}\,\mathcal{M}_{\alpha\beta\mu\nu}\delta\left(s-s'\right)\,,
\end{split}
\end{equation}
i.e. by substituting in Eq.~\eqref{eq:RumpfVariance} the average $\langle \mathcal{G}_{\alpha\beta\mu\nu} \rangle / \kappa^2$ with $\langle g_{\alpha\beta}g_{\mu\nu}\rangle = \mathcal{M}_{\alpha\beta\mu\nu}$. Indeed, we notice that by means of the multiplicative noise one easily reaches the aim of decoupling the noise and the metric correlation functions, leveraging the non-anticipating character of any functional of the metric tensor with respect to the noise. This property holds only when interpreting Eq.~\eqref{eq:SQ_mult} in the It\^o sense~\cite{Rumpf_1986}. This separation was achieved in~\cite{Rumpf_1986} by considering the internal space to the super-metric by means of super-tetrads, which can be written as
\begin{equation}
E_{\quad\alpha\beta}^{ab}=|g|^{-1/4}e_{\alpha}^{a}e_{\beta}^{b}, \quad E_{ab}^{\quad\alpha\beta}=|g|^{1/4}e_{a}^{\alpha}e_{b}^{\beta}.
\end{equation}
This operation allows to segregate the fields fluctuations in the super-tetrads while independently treating the statistical properties of the \emph{internal} noise $\eta_{ab}^{(0)}$ that in~\cite{Rumpf_1986} read
\begin{equation}
\begin{split} & \langle\eta_{\alpha\beta}\left(s\right)\eta_{\mu\nu}\left(s'\right)\rangle=\frac{2}{\kappa^{2}}\langle\mathcal{G}_{\mu\nu\alpha\beta}\rangle\delta\left(s-s'\right)\\
= & \langle\eta_{ab}^{\left(0\right)}\left(s\right)\eta_{cd}^{\left(0\right)}\left(s'\right)E_{\quad\alpha\beta}^{ab}[g]E_{\quad\mu\nu}^{cd}[g]\rangle\\
= & \langle\eta_{ab}^{\left(0\right)}\left(s\right)\eta_{cd}^{\left(0\right)}\left(s'\right)\rangle\langle E_{\quad\alpha\beta}^{ab}[g]E_{\quad\mu\nu}^{cd}[g]\rangle\\
= & 2\mathcal{G}_{abcd}^{\left(0\right)}\langle E_{\quad\alpha\beta}^{ab}[g]E_{\quad\mu\nu}^{cd}[g]\rangle\delta\left(s-s'\right),
\end{split}
\end{equation}
thus defining the variance of the noise in the internal super-space as
\begin{equation}
    \mathcal{G}_{abcd}^{\left(0\right)}=\frac{1}{\kappa^2}\mathcal{G}_{\alpha\beta\mu\nu}E_{ab}^{\quad\alpha\beta}[g]E_{cd}^{\quad\mu\nu}[g].
\end{equation}
As mentioned above, this decoupling between noise and metric fluctuations is automatically implemented already in the super-space when considering multiplicative noise as we do in the present work. For the sake of the comparison with~\cite{Rumpf_1986} we report the expression of the noise in the internal super-space
\begin{equation}
    \eta_{ab}^{\left(0\right)}=|g|^{1/4}\eta_{ab}e^{\imath\frac{\gamma}{2}}\sqrt{2\Lambda}\, \tilde{\eta}.
    \label{eq:ss-noise}
\end{equation}
In~\cite{Rumpf_1986} it is shown that the noise- and metric-fluctuations decoupling hinges on adopting the It\^o interpretation for the stochastic differential equation at the foundation of the SQ approach, which we assume to be valid also for Eq.~\eqref{eq:SQ_mult}.\\

Let us now define the action functional $S$ more precisely as
\begin{equation}
    S=\frac{1}{2\kappa}\int\text{d}^{4}x\sqrt{-g}R+\int\text{d}^{4}x\sqrt{-g}\mathcal{L}_{\text{M}}\,,
    \label{eq:S}
\end{equation}
where the first term represents the usual Einstein-Hilbert action with $\kappa = 8\pi G/c^3$, $G$ denoting the Newton's constant and $c$ the speed of light, while the second term contains the Lagrangian density of ``matter'' fields $\mathcal{L}_{\text{M}}$. As mentioned above, the parameter $\lambda$ appearing in the supermetric plays an important role, in particular, for $\lambda = -1$ the metric tensor is harmonic in superspace and the DeWitt path-integral measure is recovered~\cite{Rumpf_1986, hawking1980general} --- about the sampling of fields configurations in general relativity using a path-integral approach see also Refs.~\cite{misner1957feynman, klauder1961covariant,dewitt1962quantization,fujikawa1984path}. Moreover, for this choice the Langevin equations read
\begin{equation}
\frac{\partial g_{\mu\nu}}{\partial s}=-2\imath\left[R_{\mu\nu}-\kappa\left(T_{\mu\nu}-\frac{1}{2}g_{\mu\nu}T\right)\right]+g_{\mu\nu}e^{\imath\frac{\gamma}{2}}\sqrt{2\Lambda}\, \tilde{\eta},
\label{rula}
\end{equation}
where the term in square brackets can be identified with Einstein's equations. When matter is absent, Eq.~\eqref{rula} assumes the form of a generalized complex Ricci flow with multiplicative noise. We adopt such a choice in order to adhere as closely as possible to the original Ricci flow, representing a classical geometric flow naturally breaking diffeomorphisms covariance. Thus this amounts to dress the noise term with the geometric and physical meaning of a scalar quantity under general coordinates transformations. This new dynamics would indeed allow to study the fluctuations of the gravitational field around the saddle point $\delta S=0$, where the symmetries are implemented by the dynamics dictated by the Einstein equations. A few works in the literature have approached the Ricci-flow as a tool of analysis for GR both classically~\cite{Graf_2007} and, more recently, from the quantum perspective~\cite{frenkel2020topological}. However, the present connection to the SQ approach represents a novelty in the literature.

Finally, as reported in Eq.~\eqref{rula} the insertion of matter within this scheme happens pretty naturally and it can also be interpreted from the perspective of dynamic systems as adding a \emph{target} curvature tensor 
\begin{equation}
R_{\mu\nu}^{T}=\kappa\left[T_{\mu\nu}-\frac{1}{2}g_{\mu\nu}T\right]
\label{RM}
\end{equation}
that is used to ``shift'' the fixed point of the classical flow, thus including the presence of matter.

\section{Cosmological constant and stochastic time } \label{SSFP}
\noindent 
The Langevin equation we have introduced in \eqref{rula} enables to cast, according to the It\^o calculus  \cite{gardiner2004handbook}, an associated equation for the evolution of the probability distribution $p=p[g_{\mu \nu}(s);s]$, i.e. the Fokker-Planck equation, which reads 
\begin{equation}
\frac{\partial p}{\partial s}=-\frac{\delta}{\delta g_{\mu\nu}}\left[\imath\mathcal{G}_{\mu\nu\alpha\beta}\frac{\delta S}{\delta g_{\alpha\beta}}\,p\right]+\frac{\delta^{2}}{\delta g_{\mu\nu}^{2}}\left[g_{\mu\nu}^{2}e^{\imath\gamma}\Lambda\,p\right]\,.
\end{equation}
The steady-state solution $\partial p/\partial s=0$ can then be straightforwardly solved, providing, for an arbitrary integration constant $D$, the approximated expression
\begin{equation} \label{stesta}
p\simeq\frac{D}{g_{\mu\nu}^{2}}\exp\left[\imath\int^{g_{\mu\nu}}\mathcal{D}g_{\alpha\beta}\frac{\mathcal{G}_{\alpha\beta\rho\sigma}\frac{\delta S}{\delta g_{\rho\sigma}}}{e^{\imath\gamma}\Lambda_{0}g_{\alpha\beta}^{2}}\right]\,.
\end{equation}

We may then further inspect the configurations that extremize the steady-state solution of the Fokker-Planck distribution, imposing the first variation in the metric field of \eqref{stesta} to vanish. Since we are looking for those solutions that extremize the Fokker-Planck probability distribution and realize a saddle point in $p$, the second variation of \eqref{stesta} should be negative. \\

\subsection{Cosmological constant and stochastic noise} 
\noindent
A straightforward calculation shows that the first variation of the Fokker-Planck probability distribution vanishes when the following differential equation is fulfilled 
\begin{equation}\label{ados}
 \mathcal{G}_{\rho\sigma\mu\nu}\frac{\delta S}{\delta g_{\rho\sigma}}+2\imath e^{\imath\gamma}\Lambda g_{\mu\nu}=0\,.  
\end{equation}
To gain intuition on \eqref{ados}, we specify $S$ to be the Einstein-Hilbert action of gravity for Lorentzian manifolds plus the action of matter fields (see Eq.~\eqref{eq:S}), thus obtaining
\begin{equation}
    R_{\mu\nu}-\kappa\left(T_{\mu\nu}-\frac{1}{2}g_{\mu\nu}T\right)-\imath e^{\imath\gamma}\Lambda g_{\mu\nu}=0.
    \label{adriosca}
\end{equation}
One can recover the Einstein's equations with a real cosmological constant $\Lambda$ by setting $\gamma = \pi/2$, or otherwise just adsorbing the imaginary unit through the phase shift $\gamma\to\gamma + \pi/2$. Furthermore, the real components of the Hessian matrix of $p$ are found to be negative once evaluated on Eqs.~\eqref{adriosca}, which ensures the solutions to these latter to maximize the Fokker-Planck probability distribution $p$, hence corresponding to the classical solutions of the theory of gravity taken into account.

\subsection{Stochastic time} \label{time}
\noindent 
The choice of the quantity representing the stochastic time $s$ shall be carefully pondered. While most studies applying SQ to various field theories do not always match on the physical meaning of the stochastic time, it seems reasonable that the geometric nature of GR would require some properties to be fulfilled. As an example, one might expect the stochastic time $s$ to be represented as an affine parameter whose integral should transform as a scalar under diffeomorphisms. It has been already observed in previous works on SQ of two-dimensional causal dynamic triangulations (CDT)~\cite{Ambj_rn_2009} that the proper time $\tau$ naturally plays the role of the stochastic time. Indeed, $\tau$ possesses all the transformation properties we discussed above. However, two-dimensional dynamic triangulations are somewhat an oversimplified model as we are about to argue. In a four-dimensional theory the proper-time dependence is both explicit and encoded in the coordinate dependence. 

Let us now focus on the thermal-time-dynamics close to equilibrium, having introduced the dependence on $s$ as a further dependence in the fields. Hence, one would look for an operator like
\begin{equation}
\frac{\mbox{d}}{\mbox{d} s} = \frac{\partial}{\partial s} + \frac{\mbox{d}x^\mu}{\mbox{d} s}\nabla_\mu\,. 
\end{equation}
This represents a conformal transformation along the normal that can be applied to the metric tensor if one adds a projective component to the connection. Its singular points can be interpreted as manifold surgeries that change the topology. This observation is relevant to provide an intuitive meaning to the geometric (Ricci) gradient flow. For the stochastic version of the geometric gradient flow, the stochastic Ricci flow, this interpretation holds close to equilibrium. We also point out the link between the notion of proper time here we make reference to, and the one introduced by York \cite{York_1972}, and the connection with the formulation provided by Nambu for quantization~\cite{Nambu_1950}.\\

We may inspect now the relation between the thermal time $s$ and the proper time $\tau$. Close to equilibrium, where average quantities are constant, the connection between $s$ and $\tau$ can be determined from the identification of $\ell(s)\,\mbox{d}x^\mu/\mbox{d}s = n^\mu$ (with $\ell(s)$ having the dimensions of a length), which is also defined as the covariant derivative of the scalar time field $n_\mu = -N\nabla_\mu t$, namely 
\begin{equation}
\ell\left(s\right)\frac{\mbox{d}x^{\mu}}{\mbox{d}s}=-g^{\mu\alpha}N\partial_{\alpha}t\,,
\end{equation}
\begin{equation}
g_{\mu\nu}n^{\mu}n^{\nu}=g_{\mu\nu}\ell^{2}\left(s\right)\frac{\mbox{d}x^{\nu}}{\mbox{d}s}\frac{\mbox{d}x^{\mu}}{\mbox{d}s}=-g_{\mu\nu}\ell\left(s\right)\frac{\mbox{d}x^{\nu}}{\mbox{d}s}g^{\mu\alpha}N\partial_{\alpha}t\,,
\end{equation}
\begin{equation}
g_{\mu\nu}n^{\mu}n^{\nu}=-\ell\left(s\right)\frac{\mbox{d}t}{\mbox{d}s}N=\varepsilon(s)\,,
\end{equation}
from which it follows that
\begin{equation}
\varepsilon(s)\frac{\delta s}{\ell\left(s\right)}=-N\delta t\,. 
\end{equation}
Here, we are considering a normalization which depends on the stochastic time $s$, that behaves as an affine parameter. If one has in the limit
\begin{equation}
\lim_{s\to\infty}\varepsilon(s)\frac{\delta s}{\ell\left(s\right)}=\varepsilon\frac{\delta s}{\ell}=-N\delta t\,,
\end{equation}
then this, together with $\varepsilon = -1$, would correspond to the usual definition of the proper time
\begin{equation}
\delta \tau = N\delta t\,, 
\end{equation}
so that
\begin{equation}
s \to \ell \tau \,,
\end{equation}
with the Jacobian between $s$ and $\tau$ provided by
\begin{equation}
\frac{\mbox{d}\tau}{\mbox{d}s} = -\frac{\varepsilon(s)}{\ell(s)} \,.  
\end{equation}
Hence, in order to define a stochastic time $s$ that tends to the proper time $\tau$ at equilibrium, we need to take into account a variable normalization of the normal vector to the space-like hypersurfaces. It is possible to write the average relation as the long-time limit of an equal-time correlator
\begin{equation}
\lim_{s\to\infty}\langle n^\mu(s) n_\mu(s)\rangle = \varepsilon\,.  
\end{equation}
We can then write
\begin{equation}
\delta s=\frac{\mbox{d}s}{\mbox{d}\tau}\delta\tau=\ell(s)\sqrt{\frac{1}{c^{2}\varepsilon^{2}(s)}g_{\mu\nu}\mbox{d}x^{\mu}\mbox{d}x^{\nu}}\,, 
\end{equation}
from which one can see that the effect of a normalization of the normal vector to the space-like hypersurface is to rescale the line element by $\varepsilon^{-2}$.

\section{The Ricci flow in the Hamiltonian ADM formalism} \label{results}
\noindent
We report in this section the Hamiltonian analysis of the Ricci-flow. We start recalling the splitting of the metric in the ADM decomposition \cite{ADM} of a generic line element
\begin{eqnarray}
ds^2&=&g_{\mu \nu}\, dx^\mu dx^\nu \nonumber \\
&=& -N^2 dt^2 + h_{ij} (dx^i + N^i dt) (dx^j + N^j dt)\,,
\end{eqnarray}
in which $ h_{ij}$ is the three-metric induced on the three-dimensional spatial hyper-surfaces by the action of the projector $q^{\alpha \beta}$ on the four-dimensional metric $g_{\mu \nu}$, $N$ denotes the lapse function and finally $N^i$ stands for the shift vector. 

We define the unit time-like vector $n^\mu$, normal to the hypersurfaces of constant coordinate time $t$, to be 
\begin{eqnarray}
n_\mu=(-N, 0)\,, \qquad n^\mu=\left(\frac{1}{N}, -\frac{N^i}{N}\right)\,.
\end{eqnarray}

With these definitions, we are allowed to introduce the extrinsic curvature tensor --- this measures the curvature of the hyper-surface within the spacetime manifold, i.e.  after parallel transport respect to the space-time manifold Levi-Civita connection, the failure of a vector tangent to the hyper-surface to remain tangent to it --- observing that it rewrites    
\begin{eqnarray}
K_{ij}&=&-\nabla_{(j} n_{i)}= \\
&=& \frac{1}{2N} \left( -\partial_t h_{ij} + \!\!\!\phantom{a}^{(3)}\nabla_{(i} N_{j)} + \!\!\!\phantom{a}^{(3)}\nabla_{(j} N_{i)}  \right) \nonumber\,,
\end{eqnarray}
$\!\!\!\phantom{a}^{(3)}\nabla_i$ denoting the covariant derivative with respect to the Levi-Civita connection on the spatial hyper-surface~\footnote{As reported in Appendix~\ref{app:ext_conn}, when the extended connection $\mathcal{C}^\alpha_{\ \mu \nu}$ is taken into account, one can show that
\begin{equation}
\begin{split}\bar{\nabla}_{(\alpha}n_{\beta)} & =\nabla_{(\alpha}n_{\beta)}-\mathcal{C}_{(\alpha\beta)}^{\gamma}n_{\gamma}\\
 & =\nabla_{(\alpha}n_{\beta)}-\left(\lambda_{1}+\lambda_{2}\right)n_{\alpha}n_{\beta}\\
 & +\varepsilon\left(s\right)\left[\lambda_{3}w_{\alpha\beta}+\lambda_{4}n_{\alpha}n_{\beta}\right]\,,
\end{split}
\end{equation}
so that clearly, when projecting to the hypersurface, i.e. defining the extrinsic curvature with four-dimensional indices $\bar{K}_{\alpha\beta} = -q_{\alpha}^\mu q_{\beta}^\nu \bar{\nabla}_\mu n_\nu$, only the term proportional to $\lambda_3$ does not vanish. In this work we set for convenience $\lambda_3 = 0$ so that $\bar{K}_{ij} = K_{ij}$.}.\\

To have control over the fluctuation of the constraints, which recast the dynamics of GR at the Hamiltonian level, we provide to rewrite the Ricci flow equations in the ADM variables \cite{ADM}. We assume that the noise can be decomposed in terms of a scalar noise $\eta$, as $\eta_{\mu \nu}=\eta \, g_{\mu \nu}$.

Furthermore, we need to take into account It\^o's lemma~\cite{Ito1951,gardiner2004handbook} when considering the transformation from the metric to the ADM variables. The SQ equations transform as
\begin{equation}
\begin{split}\frac{\partial h_{ij}}{\partial s} & =\frac{\partial g_{ij}}{\partial s}\,,\\
\frac{\partial N^{k}}{\partial s} & =\frac{\partial N^{k}}{\partial g_{\mu\nu}}\frac{\partial g_{\mu\nu}}{\partial s}+\frac{\partial^{2}N^{k}}{\partial g_{\alpha\beta}\partial g_{\mu\nu}}g_{\alpha\beta}g_{\mu\nu}e^{\imath\gamma}\Lambda\,,\\
\frac{\partial N}{\partial s} & =\frac{\partial N}{\partial g_{\mu\nu}}\frac{\partial g_{\mu\nu}}{\partial s}+\frac{\partial^{2}N}{\partial g_{\alpha\beta}\partial g_{\mu\nu}}g_{\alpha\beta}g_{\mu\nu}e^{\imath\gamma}\Lambda \,.
\end{split}
\end{equation}
Because of the identity $h_{ij}=g_{ij}$, the equation for the three-metric does not transform, while for the shift and the lapse one obtains for the second derivative terms
\begin{equation}
\frac{\partial^{2}N^{k}}{\partial g_{\alpha\beta}\partial g_{\mu\nu}}g_{\alpha\beta}g_{\mu\nu}=0,\quad\frac{\partial^{2}N}{\partial g_{\alpha\beta}\partial g_{\mu\nu}}g_{\alpha\beta}g_{\mu\nu}=-\frac{1}{4}N,    
\end{equation}
where the term related to the shift vector remarkably vanishes.\\
 
An appropriate linear combination of $R_{00}$, $R_{0i}$ and $R_{ij}$ provides, after several manipulations, the RG flow of the scalar component of the Hamiltonian constraint, namely
\begin{equation} \label{rgsc}
\frac{\partial N}{\partial s}=-\frac{N}{2}\left[\frac{\imath\mathcal{H}}{\sqrt{h}}+\frac{1}{2}e^{-\imath\gamma}\Lambda+\eta\right]\,,
\end{equation}
denoting that the RG flow of the lapse function is driven by the scalar constraint $\mathcal{H}$, as it was actually intuitive to argue, but is subjected to fluctuations induced by the noise source $\eta$, while at the same time gaining what seems to be a non-zero eigenvalue at the steady state related to the square of the noise amplitude, i.e. $\Lambda$. This result points to a possible resolution of the problem of the frozen formalism~\cite{CQG_2014} by providing an effective time flow intertwined with the quantum fluctuations. At equilibrium, i.e. at the end of the relaxation, the scalar constraint is restored on average.

It is remarkable that the SQ Ricci flow equations for $N^k$ are instead not directly affected by the thermal noise that we introduced --- namely, in our assumption, $\eta_{\mu \nu}=\eta\,  g_{\mu \nu}$ --- and that the expression of the flow of the shift vector $N^k$ writes  
\begin{equation}\label{rgvc}
\frac{\partial N^{k}}{\partial s}=\frac{\imath  N\mathcal{H}^{k}}{\sqrt{h}}\,.
\end{equation}
Again, as it would have been natural to guess, the components of the vector constraint $\mathcal{H}^{k}$, which for the theory at equilibrium, namely GR, generates the space diffeomorphism transformations, is driving the flow of the components of $N^k$. At equilibrium, the diffeomorphism constraint is recovered. \\

Finally, we inspect the $ij$ components of the Ricci RG flow of the metric. These can be cast in terms of the Lie derivative $\mathcal{L}_{m}$ along the vector $m^\alpha = N \, n^\alpha$, and involve Poisson-brackets of the tri-metric with the scalar constraint $\mathcal{H}$ in the form   
\begin{equation}
\frac{\partial h_{ij}}{\partial s}=\frac{1}{N}\mathcal{L}_{m}\left[\mathcal{H},h_{ij}\right]+\left[\mathcal{H},\left[\mathcal{H},h_{ij}\right]\right]+\frac{h_{ij}\mathcal{H}}{2\sqrt{h}}-h_{ij}\eta\,,
\end{equation}
where the noise source $\eta$ is appearing again in the flow equations.

\section{Physical Interpretation of the Ricci flow equations} \label{PI}
\noindent
In this section we provide a physical interpretation of the components of the Ricci RG flow equations that have been recovered in the previous section. 

\subsection{Fluctuating away from the vector constraints: chaos and intermittency in Navier-Stokes equations} 
\noindent
The duality between the solutions of the incompressible Navier-Stokes equation in $d$-dimensions and their uniquely associated solutions of the vacuum Einstein equations in $(d + 1)$-dimensions was pointed in \cite{Bredberg_2012}, providing a rigorous realization of the holographic duality between fluids and horizons, discussed in the literature a couple of decades before the AdS/CFT correspondence.\\

Within this extended framework, it is relevant to emphasize once again that the ADM transformation of the metric tensor induces a non-trivial result for the equation of the shift vector $N^k$, Eq.~\eqref{rgvc}, i.e. the noise term vanishes. Even more interestingly, the usual expression of the conjugated momentum to the three-dimensional metric tensor is proportional to the Brown-York stress tensor $T^{ij}$
\begin{equation}
    \Pi^{ij} = \sqrt{h}(Kh^{ij} - K^{ij}) = \frac{1}{2}\sqrt{h} T^{ij}\,.
\end{equation}
As it was shown in~\cite{Bredberg_2012}, the ordinary divergence $\partial^{i} T_{ij}$ is proportional to the Navier-Stokes equations, yielding
\begin{equation}\label{nseq}
    r_{c}^{3/2}\partial^{k}T_{ki}=\partial_{t}v_{i}-\zeta\partial^{2}v_{i}+\partial_{i}P+v^{k}\partial_{k}v_{i}=0\,,
\end{equation}
where $\zeta$ is the kinematic viscosity and $v_i$ represents the velocity field of an incompressible fluid. In particular, the incompressibility condition directly follows from the divergence $r_c^{3/2}\partial_k T^{kt} = \partial_k v^k = 0$ --- see e.g. \cite{Bredberg_2012}.

Given the usual definition of the super-momentum constraint
\begin{equation}\label{supmom}
    \mathcal{H}^i = 2 \!\! \phantom{a}^{(3)}\nabla_{k} \Pi^{ki}\,,
\end{equation}
it follows that imposing the constraint $\mathcal{H}^k=0$ implies Eq.~\eqref{nseq}. 

It is relevant at this stage to point out an interesting connection with turbulence theory, which may play a role for the analysis of the stochastic background of gravitational waves over first order phase transitions (of possible dark matter candidates) in particle physics. 
It is well known that turbulent flows can develop in low-viscosity liquids~\cite{Frisch_1995} when subject to a stochastic forcing. Indeed, one can recast Eq.~\eqref{rgvc} as
\begin{equation}\label{nsforced}
    r_{c}^{3/2}\partial_{k}T^{ki}=\frac{1}{N}\frac{\partial N^{i}}{\partial s}\,.
\end{equation}
Since $N$ and $N^{k}$ (through the normalization $\varepsilon(s)$) are both stochastic fields, one recovers that Eq.~\eqref{rgvc} represents a Navier-Stokes equation with stochastic forcing, which in turn can inject into the system an intermittent noise, qualitatively different from the Gaussian noise~\cite{Frisch_1995} explicitly introduced through the term $\eta(x,s)$.\\

The multiplicative ansatz for the stochastic noise is responsible for its own disappearance from the gradient flow of the vector constraint. This ensures that at the equilibrium limit, in which the stochastic forcing disappears, the system approaches a classical (hydrodynamic) limit, without the need of performing any statistical average or requiring the noise to vanish.

\subsection{Emergence of the Kardar-Parisi-Zhang equation}
\noindent 
We finally inspect the physical meaning of the $ij$ space components of the Ricci RG flow equations, adapting those latter to spherical symmetric metrics of the form 
\begin{equation} \label{sfe}
ds^2= - N^2 dt^2 + e^{\mu(r)} dr^2 + r^2 d\theta^2 + r^2 \sin^2 \theta d\phi^2\,,      
\end{equation}
with $N^2=e^{\nu(r)}$. \\

Imposing $\mu(r)=-\nu(r)$ on the $\mu(r)$ and $\nu(r)$ functions, before performing the functional variation necessary to determine the Ricci flow equations, would amount to impose the Einstein equations, namely to consider the solutions at equilibrium. We shall therefore refrain from this substitution, and perform first the functional variation of the Einstein-Hilbert action for the spherically symmetric metrics taken into account in Eq.~\eqref{sfe}, and then impose the condition $\mu(r)=-\nu(r)$ at the very end.

We may start from the variation  
\begin{equation}
\frac{\partial g_{00}}{\partial s}=-2\imath R_{00}+g_{00}\, e^{\imath\frac{\gamma}{2}}\sqrt{2\Lambda}\, \tilde{\eta}\,,
\end{equation}
which according to the It\^o rule transforms into
\begin{equation}
\begin{split}\frac{\partial\nu}{\partial s}= & \frac{\text{d}\nu}{\text{d}g_{00}}\frac{\partial g_{00}}{\partial s}+e^{\imath\gamma}\Lambda\,g_{00}^{2}\frac{\text{d}^{2}\nu}{\text{d}g_{00}^{2}}\,,\\
\frac{\partial\nu}{\partial s}= & -\imath\left[\frac{\text{d}^{2}\nu}{\text{d}r^{2}}+\frac{2}{r}\frac{\text{d}\nu}{\text{d}r}+\left(\frac{\text{d}\nu}{\text{d}r}\right)^{2}\right]e^{\nu}\\
 & -\frac{1}{2}e^{\imath\frac{\gamma}{2}}\sqrt{2\Lambda}\,\left(\tilde{\eta}+e^{\imath\frac{\gamma}{2}}\sqrt{2\Lambda}\right)\,.
\end{split}
\label{KPZ}
\end{equation}
We observe that, at the right hand-side of Eq.~\eqref{KPZ}, the quantity $e^{\nu} \rightarrow 1$ for $r\!>\!\!> \!r_S$, where $r_S=2Gm$ denotes the Schwarzschild radius of a black hole of mass $m$. Based on these observations, we can conclude that Eq.~\eqref{KPZ} coincides with the Kardar-Parisi-Zhang (KPZ) equation~\cite{Kardar_1986} in spherical coordinates, considering $\Lambda$ as negligible, which is a good approximation even reasonably far from the event horizon. We notice finally that, in achieving this result, it has been crucial to switch to the $\mu$ and $\nu$ coordinates that allow the noise to become additive, as indeed in the case of the KPZ equation.\\

We have then recovered that, close to equilibrium (at the asymptotic limit of spherically symmetric metrics) the (Ricci) RG flow of gravity is described by the non-trivial properties of the probability distribution of the KPZ universality class. We further notice that, in a one-dimensional space, the KPZ equation is linked to the Burgers equation. This latter provides the simplest model yielding intermittent fluctuations, and hence intermittent statistics, similarly to the Navier-Stokes statistics for turbulent flows. \\

Addressing the profound consequences of this result, we shall emphasize that intermittency is not self-similarity, like it is usually assumed in condensed matter and quantum field theory. For instance, the Wilsonian approach to the RG flow is based on the self-similarity of the fluctuations, close to the critical point of the phase transition, this latter denoting the point where one applies the RG flow in order to compute the critical exponents and the critical amplitudes. \\ 

We further comment that, since KPZ has a general role, it is remarkable and reassuring that, at least for a certain class of metric, this equation can be recovered in our analysis. The solutions to the KPZ equation represent indeed a universality class as general as the one provided by the solutions to the Brownian motion equation. This latter instantiates a continuum scaling limit for a very large class of random processes. Its properties, including the distribution function or the regularity, have been widely studied in the literature. The KPZ universality class was proposed over the last two decades, to describe a wealth of relevant physical and probabilistic models, which include the one-dimensional interface growth processes and the interaction of systems of particle and polymers in random environments, all phenomena which retain a new statistics and display unusual novel scaling features. The elements of the KPZ universality class are solutions to a non-linear stochastic partial differential equations. For the KPZ equation, the exact one-point distribution of the solutions can be determined, with narrow wedge initial data, and remarkable connections with directed polymers in random media can be recovered.

\section{The cosmological constant in the new framework} \label{CC}
\noindent 
A paradigmatic study case is provided by the RG flow of the cosmological constant. This can be investigated adapting the Ricci flow stochastic equation to the Friedman Lema\^itre Robertson Walker (FLRW) background, which reads in co-moving spherical coordinates 
\begin{equation}
ds^2=-N^2 dt^2 + a^2(t) \left[\frac{(dr)^2}{1-k r^2} + r^2 (d\theta^2+ \sin^2 \theta d\phi^2) \right]\,,    
\end{equation}
with $k=0, +1, -1$ denoting respectively vanishing, positive and negative space curvature. Within these coordinates, 
\begin{equation}
R=6\left( \frac{\ddot{a}}{a} +\left(\frac{\dot{a}^2}{a^2} \right) +\frac{k}{a^2} \right)\,,
\end{equation}
where dot denotes the derivative with respect to the co-moving time, and $\sqrt{-g}=N a^3 r^2 \sin \theta$. The Einstein-Hilbert action casts
\begin{equation} \label{aflrw}
S=\frac{1}{2\kappa}\int\text{d}^{4}x\sqrt{-g}\left(R+2\kappa\mathcal{L_{\text{M}}}+3\varepsilon\left(s\right)\lambda_{2}^{2}\right)\,.
\end{equation}

\subsection{Ricci RG flow of the cosmological constant}
\noindent 
Varying the action in Eq.~\eqref{aflrw} with respect to the fields $a$, $N$ and $\lambda_2$, we find the components of the Einstein equations, to which we shall add the noise terms. In particular, we define the equation for $\lambda_2$ as
\begin{equation}
    \frac{\partial\lambda_{2}}{\partial s}=-\frac{1}{3}\frac{\imath}{\sqrt{-g}}\frac{\delta S}{\delta\lambda_{2}}+\lambda_{2}e^{\imath\frac{\gamma}{2}}\sqrt{2\Lambda}\tilde{\eta}
\end{equation}
thus keeping dimensional consistency.

Bearing in mind the possible relevance of the running of the cosmological constant for the resolution of the Hubble tension, determined by the mismatch between the measurements of the Hubble parameter at redshift $z\simeq 1100$ and $z=1 \div 2$, we disregard over the cosmological epoch under scrutiny the running of Newton’s constant $G\propto \kappa$.\\

After some manipulations, we recover the system for the stochastic differential equations
\begin{equation}
\begin{split}\frac{\partial a}{\partial s}&=-\imath\frac{a}{3N^{2}}\Bigg[\frac{6kN^{2}}{a^{2}}+3\left(\dot{H}+3H^{2}\right)-N^{2}g^{ij}R_{ij}^{T}\\&\qquad\qquad-\frac{9}{2}N^{2}\lambda_{2}^{2}-\frac{5}{4}N^{2}e^{\imath\left(\gamma+\frac{\pi}{2}\right)}\Lambda\Bigg]+\frac{1}{2}ae^{\imath\frac{\gamma}{2}}\sqrt{2\Lambda}\tilde{\eta}\,,\\
\frac{\partial N}{\partial s}&=\imath\left[-\frac{3}{N}\left(\dot{H}+H^{2}\right)\!+\!\frac{1}{N}R_{00}^{T}\!+\!\frac{3N}{2}\lambda_{2}^{2}\!+\!\frac{N}{4}e^{\imath\left(\gamma+\frac{\pi}{2}\right)}\Lambda\right]\\&+\frac{N}{2}e^{\imath\frac{\gamma}{2}}\sqrt{2\Lambda}\tilde{\eta}\,,\\
\frac{\partial\lambda_{2}}{\partial s}&=-\imath\left[\frac{1}{\kappa}\varepsilon\left(s\right)+e^{\frac{\imath}{2}\left(\gamma+\pi\right)}\sqrt{2\Lambda}\tilde{\eta}\right]\lambda_{2}\,,
\end{split}
\end{equation}
where $H=\dot{a}/a$ denotes the Hubble function, and we have implemented the It\^o variable transformations for $a$ and $N$, which read respectively 
    \begin{equation}
        \frac{\partial a}{\partial s}=\frac{\partial a}{\partial g_{ij}}\frac{\partial g_{ij}}{\partial s}+\frac{\partial^{2}a}{\partial g_{ij}^{2}}g_{ij}^{2}e^{\imath\gamma}\Lambda\,,
    \end{equation}
  and 
    \begin{equation}
        \frac{\partial N}{\partial s}=\frac{\partial N}{\partial g_{00}}\frac{\partial g_{00}}{\partial s}+\frac{\partial^{2}N}{\partial g_{00}^{2}}g_{00}^{2}e^{\imath\gamma}\Lambda\,,
    \end{equation}
and used the fact that for the second derivatives it holds:
    \begin{equation}
\frac{\text{d}^{2}a}{\text{d}h_{ij}^{2}}h_{ij}^{2}=-\frac{5}{12}a,\qquad\frac{\partial^{2}N}{\partial g_{00}^{2}}g_{00}^{2}=-\frac{1}{4}N\,.
\end{equation}

The detailed analysis of the system of stochastic differential equations for $a$, $N$ and $\lambda_2$ will be presented elsewhere \cite{LuMaVi}. Nonetheless, we are already in the position to derive some preliminary conclusions about the running of the cosmological constant by solving the stochastic differential equation for $\lambda_2$: this equation is that of a complex harmonic oscillator with stochastic frequency. Following Ref.~\cite{gardiner2004handbook}, we do not provide immediately the interpretation of the equation for $\lambda_2$ in the Ito sense, but rather in the Stratonovich way. Changing the last equation to the Ito interpretation we then derive
\begin{equation}
\begin{split}\text{d}\lambda_{2} & =\left[\imath\left(-\frac{1}{\kappa}\varepsilon+\Lambda\sin\gamma\right)+\Lambda\cos\gamma\right]\lambda_{2}\, \text{d}s\\
 & +\lambda_{2}\, e^{\imath\frac{\gamma}{2}}\sqrt{2\Lambda}\, \text{d}W.
\end{split}
\end{equation}
Solving by adopting the It\^o calculus, one obtains for the two point correlation $\langle\lambda_{2}\left(s\right)\lambda_{2}^{*}\left(s'\right)\rangle$ in the $s'\to s$ indeed: 
\begin{equation}
\langle|\lambda_{2}\left(s\right)|^{2}\rangle=\langle|\lambda_{2}\left(0\right)|^{2}\rangle\exp\left\{ 4\Lambda \cos\gamma\,s\right\} .
\label{ccs}
    \end{equation}
Being the cosmological time oriented in a similar way than the thermal time, this provides a cosmological constant the value of which increases exponentially, as time increases, but with a time-constant that is supposed to be big enough to maintain a relatively small variation over cosmological times. In a simplified (still unrealistic) framework in which any matter and radiation contribution to the Friedmann equations is neglected (i.e. $R_{\mu\nu}^{\text{T}}=0$), the Hubble parameter coincides modulo factors with the square root of the cosmological constant, and hence runs with the stochastic time according to the same functional dependence.

\subsection{A way out from the Hubble tension?}
\noindent 
The falsification of our framework may have deep repercussions in cosmology, in particular in the estimate of the Hubble constant, measuring the current (accelerated) expansion of the Universe.\\

Observations announced in 1998 of distance–redshift relation for Type Ia supernovae~\cite{Perlmutter_1998} indicated that the Universe is currently undergoing an accelerated expansion --- for a recent analysis see Ref.~\cite{Riess:2021jrx}. When combined with measurements of the cosmic microwave background radiation these implied a value of $\Omega_\Lambda \simeq 0.7$~\cite{Baker_1999}, a result which has been supported and refined by more recent measurements deploying CMB observables~\cite{Plank2015_results, Planck:2018vyg}.\\

Several different origins have been advocated to account for an accelerating Universe. The cosmological constant is in most respects the simplest solution. The current standard model of cosmology, the $\Lambda$-CDM model, includes the cosmological constant, which is measured to be on the order of $10^{-52}$ ${\rm m^{-2}}$ (it is often expressed as $10^{-35}$  ${\rm s^{-2}}$ or as $10^{-122}$ by multiplication with then square Planck length, i.e. $10^{-70}$ ${\rm m^2 }$). This value is based on recent estimations of vacuum energy density: $\rho_{\rm vac} = 5.96 \times 10^{-27} {\rm kg/m^3}$~\cite{Planck2015_cosmological}.\\

The two aforementioned measurements, respectively the ones achieved by the redshift relation for Type Ia supernovae and by the CMB measurements, provided different values for the Hubble constant, and hence for the cosmological constant. The SH0ES experiment is providing regular updates of the astronomical measurement, all of them within the same range of ever narrowing error. The most recent update, in 2019, fixed the value of the Hubble constant to be $74.03 \pm 1.42$ (in kilometres per second per $3.26$ million light-years). Concerning the CMB measurements, the ESA Planck satellite release that came in 2014 fixed the value of Hubble constant to be $67.4 \pm 1.4$ (in kilometres per second per $3.26$ million light-years).\\ 

We conjecture that this observed gap between the two values, of about 10\%, could be explained resorting to a mild cosmological-time variation of the cosmological constant, which is induced by its Ricci RG flow, as described in Eq.~\eqref{ccs}. This hypothesis, which suggests that the variation of the mismatch in the measurements of the cosmological constant is due to the quantum fluctuations of space-time, deserves a more detailed analysis that we postpone to a forthcoming study.

\section{Conclusions and outlooks} \label{CO}
\noindent 
We have investigated the physical richness of the Ricci flow, clarifying how the stochastic quantization to it inspired can provide the renormalization group flow of theories of gravity. We have commented on the deep physical meaning of the Ricci flow cast within the Hamiltonian formulation in ADM variables, unveiling that the fluctuations of the metric tensor components are forced by the renormalization group flow to fulfil the Kardar-Parisi-Zhang equation~\cite{Kardar_1986}, characterized by non-trivial fluctuations. We have discussed the appearance of chaos and intermittency, due to the mapping of the equations for the shift vector to the Navier-Stokes equations.

A remarkable by-product of the stochastic quantization of gravity \`a la Ricci flow here introduced is the emergence, as a macroscopic effect of quantum geometry, of the cosmological constant as the square of the amplitude of the multiplicative noise considered in the Langevin equation. Once the stationary solution is recovered, the Fokker-Planck probability distribution can be shown to be dominated by configurations that fulfil the Einstein equation with a cosmological constant that depends on the amplitude of the noise. This is a novel feature provided by our approach, which sheds light in an unprecedented way to the problem of the cosmological constant.

The imposition of the scalar Hamiltonian constraint at equilibrium, at the fixed point of the stochastic Ricci (RG) flow, realizes the time-reparametrization invariance of GR. Removing this symmetry due to the quantum fluctuations of the metric tensor (and of the matter fields), technically imposed by the random noise sources, enables time de-parametrization in an unprecedented and hitherto unexplored way. 

Within this context, the non-relativistic limit of Eq.~\eqref{rgsc} can instantiate a gravitationally-induced collapse of the wave function, as some of us have proposed in Ref.~\cite{lulli2023stochasticricciflowdynamics}. The collapse is recovered at the equilibrium, where the Hamiltonian scalar constraint is imposed on the wave-functional of the gravitational and matter fields. Recovering a meaningful relativistic formulation in any possible model able to describe the collapse of the wave-function is a longstanding challenge. The measurement problem in quantum mechanics is a subtle topic of research, which is not our intention to cover about here, and for which we rather refer to recent textbooks \cite{jordan2024quantum,wiseman2010quantum} and detailed experimental studies --- see e.g. Ref.~\cite{Minev_2019}. The novel perspective on the measurement problem proposed in Ref.~\cite{lulli2023stochasticricciflowdynamics} is inspired by the statistical mechanics description of large N-body systems. The measurement problem has been then reformulated within the context of the stochastic quantization of fields, deriving the master equation for the collapse of the wave-function from the Langevin equation of the system.

Our considerations here are also reminiscent of a way of understanding gravity that is analogical to several notable systems in soft-condensed matter. These systems undergo ``yielding transition'' that possess several features typical of critical phenomena. We can then separately assume that large N-body systems may fluctuate around configurations of ``equilibrium'' --- in which the (gravitational) Hamiltonian constraints are implemented --- and that these fluctuations can be actually modeled, in the semi-classical limit, resorting to the Ricci RG flow description. We also observe that the Ricci RG flow can be re-expressed in terms of the Hamiltonian variables, and it makes contact, in the semi-classical limit, with the Wheeler-DeWitt equation. A description in the Hamiltonian variables either of the Ricci flow, or of its version relaxing the metric tensor towards the GR solutions --- namely the flow generated by the tensor corresponding to the difference between the Ricci tensor and the Ricci target in Eq.~\eqref{RM} --- would encode the use of the Hamiltonian of the system, de-parametrized at equilibrium through the introduction of the thermal time. This can finally allow to recover in the semiclassical limit the Schr\"odinger equation.

This framework then suggests that the Ricci flow can be seen as a dynamical equation that:~i) yields the Schr\"odinger equation at the fixed point;~ii) provides away from the equilibrium, around the fixed point, the dynamical description (in the thermal time of the Ricci flow) of the ``flux" of the wave-function during its relaxation towards the ``eigenstate" individuated by the measurement process. In this sense, the thermal time can parametrize the RG flow of the energy scale of the interaction involved (system-apparatus interaction) in the localization process. In other words, the time, space and energy scales involved depend on the details of the matter interaction, which enter the matter Ricci-target --- see e.g. Eq.~\eqref{RM} --- in the general relativistic version of the Ricci flow. The proof of this conjecture and the implementation of this mechanism for the geometric collapse of the wave-functions were provided in Ref.~\cite{lulli2023stochasticricciflowdynamics}.

The renormalization group flow scheme that we have deepened through this study can be extended to the Wilsonian non-perturbative attempt for the quantization of gravity, can be implemented in an asymptotic safety scenario, and further adopted in order to describe the gravitational collapse of the wave-function, and to shed light onto the measurement problem in quantum field theory and quantum mechanics. As an immediate application of this framework, the Hubble tension recently observed between cosmological and astronomical measurements of the cosmological constant has been conjectured to be possibly solved, the mismatch between the measurements emerging as a macroscopic manifestation of quantum gravity, due to the stochastic geometric quantum fluctuations regulated by the geometric (Ricci) renormalization group flow. According to the latest experiments, the Hubble tension is still present \cite{Riess:2024ohe}, with about 5-6 $\sigma$ discrepancy, despite improved measurements. The disagreement between early (CMB-based) and late (distance ladder) methods could certainly still be explained in terms of possible ``systematic errors'' in distance calibrations (though JWST has reduced some uncertainties \cite{Riess:2024ohe}).  Nonetheless, most part of the community is considering since years that this discrepancy actually suggests that ``new physics beyond $\Lambda$CDM'' is required, involving several different scenarios, including early dark energy,  evolving dark matter/dark energy, modified gravity, neutrino interactions. Our proposal of explanation of the Hubble tension relies on the running of the cosmological constant according to the stochastic geometry (Ricci) flow. This is an effect that would fall into the class of modified gravity models, and would still be defined to happen in the infrared limit of the theory \footnote{This mechanism we are developing to explain the Hubble tension concerns cosmological times not antecedent the recombination epoch, when the running of the Newton constant can be considered not to be relevant.} In a forthcoming analysis we will provide a quantitative prediction of the value of the cosmological constant, taking into account matter and radiation within a stochastic framework, as well the different cosmological ages of the universe.

\begin{acknowledgments}
\noindent 
We wish to thank A.~Addazi, A.~Bassi, C.~Curceanu, C.~Fields, U.~Moschella, G.~Parisi, R.~Pasechnik, K.~Piscicchia, M.~Ramsey-Musolf, M.~Reichert, M.~Sakellariadou, L.~Visinelli and Y.S.~Wu for inspiring discussions over the course of the investigation that lead to this draft. 
A.M.~wishes to acknowledge support by the NSFC, through the grant No. 11875113, the Shanghai Municipality, through the grant No.~KBH1512299, and by Fudan University, through the grant No.~JJH1512105. M.L. and X.S. wish to acknowledge support by NSFC grant No.~12050410244.
\end{acknowledgments}

\appendix
\section{Ricci flow and non-linear quantum field theory}
\noindent
Now we provide an intuitive link between the Ricci flow and the RG flow of a non-linear quantum field theory. At this purpose, it is sufficient to recall the inspection of the bosonic non-linear sigma models \cite{Fri1, Fri2} that map a two-dimensional spacetime (world-sheet) into a curved Riemannian $d$-dimensional target manifold. The Nambu-Goto action of this theory casts, without accounting for a boundary term,
\begin{equation} \label{NGa}
S=\alpha' \int_{\mathcal{M}} d^2\sigma \sqrt{h}   h^{ab}(\sigma ) g_{ij}(X) \frac{\partial X^i}{\partial \sigma^a} \frac{\partial X^j}{\partial \sigma^b} \,,
\end{equation}
with $\alpha'$ a constant with the dimension of the square of a distance (``string scale''), $\mathcal{M}$ a two-dimensional  world-sheet spacetime with metric $h_{ab}$, $\sigma=(\sigma^1,\sigma^2)$ coordinates on the world-sheet and $X^i$ coordinates on the target $d$-dimensional manifold that are valued on the worldsheet. The action in Eq.~\eqref{NGa} can be then thought to introduce a theory of $d$ coupled scalar fields provided with a non-canonical kinetic term and coupling constants specified by $g_{ij}(X)$.

For this theory, remarkably, the RG flow can be expressed perturbatively in a power series of $\alpha'$, finding \cite{Fri1, Fri2}
\begin{equation}
\frac{\partial g_{ij}}{\partial \lambda} = -\alpha' R_{ij} -\frac{{\alpha'}^2}{2} R_{iklm} \,  R_j^{\ klm} + \cdots   \,.
\end{equation}
This allows to recover, up to first order in $\alpha'$, the Ricci tensor $R_{ij}$ as the generator of the RG flow for the coupling constant $g_{ij}$.

On the other side, the Ricci flow provides a geometrically intuitive insight to unveil the fate of the conformal symmetry at the quantum level, the emergence of anomalies, and the very renormalization group (RG) flow of the theory.

\section{Torsion and the cosmological constant}\label{app:ext_conn}
\noindent
We may now inspect the emergence of the term that at equilibrium, when the Einstein equations hold, provides the cosmological constant. Not surprisingly, this term can be again related, through the extension of the gravitational connection so as to encode non-torsional components, to the inclusion of matter. This is a link that becomes manifest within the first order formulation of the Einstein-Hilbert action of gravity, when the torsional parts of the connection are integrated out, being recast thanks to the Cartan structure equations in terms of the (fermionic) matter fields. We show that these additional components of the gravitational connection are ``conformal'', namely are connected to a conformal transformation of the metric. We keep focusing in this section on the second order formalism, without seeking for an interpretation of the conformal degree of freedom in term of matter. \\ 

Moving from the Christoffel connection $\Gamma_{\alpha\beta}^{\gamma}$, we first define the extended connection $\bar{\Gamma}_{\alpha\beta}^{\gamma}$ by addition of a non-symmetric component $\mathcal{C}_{\alpha\beta}^{\gamma}$, namely 
\begin{equation}
\bar{\Gamma}_{\alpha\beta}^{\gamma}=\Gamma_{\alpha\beta}^{\gamma}+\mathcal{C}_{\alpha\beta}^{\gamma}\,.
\end{equation}
The extended covariant derivative then casts 
\begin{equation}
\begin{split}\bar{\nabla}_{\mu}V^{\alpha} & =\partial_{\mu}V^{\alpha}+\Gamma_{\mu\beta}^{\alpha}V^{\beta}+\mathcal{C}_{\mu\beta}^{\alpha}V^{\beta} \,, \\
\bar{\nabla}_{\mu}V^{\alpha} & =\nabla_{\mu}V^{\alpha}+\mathcal{C}_{\mu\beta}^{\alpha}V^{\beta}\,,
\end{split}
\end{equation}
respectively for a covariant and contravariant vector. The covariant derivative of the metric tensor then reads  $\bar{\nabla}_{\mu}g_{\alpha\beta}=-2\mathcal{C}_{\mu(\alpha}^{\gamma}g_{\beta)\gamma}$. \\

We can then consider the possible parametrization of the extended part of the connection
\begin{equation}
\mathcal{C}_{\alpha\beta}^{\gamma}=\lambda_{1}\delta_{\alpha}^{\gamma}u_{\beta}+\lambda_{2}u_{\alpha}\delta_{\beta}^{\gamma}+\lambda_{3}w_{\alpha\beta}u^{\gamma}+\lambda_{4}u_{\alpha}u_{\beta}u^{\gamma}\,.
\end{equation}
The parametrization of $\mathcal{C}_{\alpha\beta}^{\gamma}$  adopts here the normal $u_\alpha$ to the three-dimensional hypersurfaces in the ADM decomposition of $g_{\mu \nu}$, and fixes $w_{\alpha \beta}$ as the projector on the three-dimensional hypersurfaces.  \\

The extended Einstein-Hilbert action can be derived starting from the extended Riemann tensor
\begin{equation}
\bar{R}_{\;\beta\gamma\delta}^{\alpha} =\partial_{\gamma}\bar{\Gamma}_{\beta\delta}^{\alpha}-\partial_{\delta}\bar{\Gamma}_{\beta\gamma}^{\alpha}+\bar{\Gamma}_{\beta\delta}^{\mu}\bar{\Gamma}_{\mu\gamma}^{\alpha}-\bar{\Gamma}_{\beta\gamma}^{\mu}\bar{\Gamma}_{\mu\delta}^{\alpha}\,,
\end{equation}
which recasts as
\begin{equation}
\bar{R}_{\;\beta\gamma\delta}^{\alpha}=R_{\;\beta\gamma\delta}^{\alpha}+\mathcal{C}_{\beta\delta}^{\mu}\mathcal{C}_{\mu\gamma}^{\alpha}-\mathcal{C}_{\beta\gamma}^{\mu}\mathcal{C}_{\mu\delta}^{\alpha}+\nabla_{\gamma}\mathcal{C}_{\beta\delta}^{\alpha}-\nabla_{\delta}\mathcal{C}_{\beta\gamma}^{\alpha}\,,
\end{equation}
and then contracting first once with the metric to find the Ricci tensor
\begin{equation}
\begin{split}\bar{R}_{\beta\delta} & =\bar{R}_{\;\beta\alpha\delta}^{\alpha}=R_{\beta\delta}+\mathcal{C}_{\beta\delta}^{\mu}\mathcal{C}_{\mu\alpha}^{\alpha}-\mathcal{C}_{\beta\alpha}^{\mu}\mathcal{C}_{\mu\delta}^{\alpha}+\nabla_{\alpha}\mathcal{C}_{\beta\delta}^{\alpha}-\nabla_{\delta}\mathcal{C}_{\beta\alpha}^{\alpha}\end{split}\,,
\end{equation}
and a second time with the metric tensor to recover the Ricci scalar
\begin{equation}
\begin{split}\bar{R} & =g^{\beta\delta}\bar{R}_{\beta\delta}=R+g^{\beta\delta}\left(\mathcal{C}_{\beta\delta}^{\mu}\mathcal{C}_{\mu\alpha}^{\alpha}-\mathcal{C}_{\beta\alpha}^{\mu}\mathcal{C}_{\mu\delta}^{\alpha}\right)\\
& \ \ \  +g^{\beta\delta}\left(\nabla_{\alpha}\mathcal{C}_{\beta\delta}^{\alpha}-\nabla_{\delta}\mathcal{C}_{\beta\alpha}^{\alpha}\right)\,.
\end{split}
\end{equation}
The Einstein-Hilbert Lagrangian density finally reads
\begin{equation} \label{eeq}
\begin{split}
\sqrt{-g}\bar{R} & =\sqrt{-g}R+\sqrt{-g}g^{\beta\delta}\left(\mathcal{C}_{\beta\delta}^{\mu}\mathcal{C}_{\mu\alpha}^{\alpha}-\mathcal{C}_{\beta\alpha}^{\mu}\mathcal{C}_{\mu\delta}^{\alpha}\right) \\
& \ \ +\partial_{\alpha}\left[\sqrt{-g}\left(g^{\beta\delta}\mathcal{C}_{\beta\delta}^{\alpha}-g^{\alpha\beta}\mathcal{C}_{\beta\delta}^{\delta}\right)\right]\,,
\end{split}
\end{equation}
in which the last term in Eq.~\eqref{eeq} is a boundary term that does not affect the Einstein equations of motion. \\

We finally examine the extended component of the action, i.e. the second term of Eq.~\eqref{eeq}, within the round brackets, by substitution of the expression of the $\mathcal{C}_{\mu\alpha}^{\alpha}$ connection, we find 
\begin{eqnarray} \label{lam}
&&\sqrt{-g}g^{\beta\delta}\left(\mathcal{C}_{\beta\delta}^{\mu}\mathcal{C}_{\mu\alpha}^{\alpha}-\mathcal{C}_{\beta\alpha}^{\mu}\mathcal{C}_{\mu\delta}^{\alpha}\right)\\
&&\qquad =\sqrt{-g}\left[(\lambda_{2}^{2}+\lambda_{3}^{2}) \left(D-1\right)u_{\mu}u^{\mu} \right]\,. \nonumber 
\end{eqnarray}
Eq.~\eqref{lam} provides an expression of the cosmological constant in terms of the square of the fields entering the definition of $\mathcal{C}$. It turns out that, since $\lambda_{i} \sim \sqrt{\Lambda}$ (for $i=2,3$), the quadratic $\lambda_{i}$ fields undergo a non-local expansion in terms of the derivative of the cosmological constant $\Lambda$, which hence acquires a dynamics within this framework. In other words, according to the stochastic geometric flow we have introduced, the fields $\lambda_{i}$, even though might fluctuate around their zero-values minimizing the action, may still have under stochastic flow non-vanishing quadratic average mimicking the cosmological constant. This will be obtained by the substitution $w_{\alpha\beta}=q_{\alpha\beta}$ and $u^\mu = n^\mu$ where $q_{\alpha\beta}$ and $n^\mu$ are the projector and the normal to the equal-time hypersurface.


\section{Discussion} \label{Di}
\noindent 
In this section, we discuss several aspects related to the stochastic quantization method, the differential structure underlying the Ricci-flow considered and the physical interpretation of the observables and relevant quantities involved in the analysis we developed.

\subsection{RG flow and proper stochastic time} \label{RGST}
\noindent 
Alternative perspectives have been hitherto developed to uncover the possibility of a non-perturbative quantization of gravity. In the asymptotic safety (AS) scenario the search for a non-trivial fixed point that would enable to control the UV-limit behaviour of the coupling constant has remained an urgent question to which has been provided so far only a partial tentative answer. The development of different pathways to the quantization of gravity, including the exact renormalization group (ERG) approach, as well as CDT, allowed to unveil un-expected properties of space-time while approaching the Planck scale, and hence unveil the behaviour of its spectral dimension $d_s$. This latter has been found to encode a fractal behaviour \cite{FB1,FB2}, due to the emergence of a ground state spacetime acquiring at large scales the classical vale $d_s=4$, and at short scales $d_s=2$. The scale-dependence of the fractal dimension is generally present also for a spacetime with a quantum group of symmetry, hence showing similarities with other quantum gravity approaches \cite{Benedetti:2008gu}.\\

The investigation of the fractal properties of spacetime is intimately connected to the thermal nature of the diffusion time. As a matter of fact, the spectral dimension is defined for a generic Riemannian (or pseudo-Riemannian) manifold $\mathcal{M}$, with possibly curved metric $g_{\mu \nu}$, as
\begin{equation}\label{hke}
\partial_s K(x,y;s)+ \Delta_x K(x,y;s)=0\,,
\end{equation}
to be solved with the initial condition $K(x,y;s)=\delta(x-y)/\sqrt{\pm g}$ (the sign $\pm$ accounting the Riemannian or pseudo-Riemannian signature of the manifold $\mathcal{M}$), having denoted the Laplacian as $\Delta_x=g^{\mu\nu} \nabla_{\mu}\nabla_{\nu}$, with $\nabla_{\mu}$ covariant derivative, and $K(x,y;s)$ the heat kernel on $\mathcal{M} \times \mathcal{M} \times \mathbb{R}_+$.

The spectral dimension is related to the trace of the heat kernel by the relation 
\begin{equation}
d_s\equiv -2 \frac{\partial {\rm Tr}K }{\partial \log s} \,,
\end{equation}
in which ${\rm Tr}K$ undergoes the expansion in the heat-kernel time \cite{Vassi}
\begin{equation}
{\rm Tr}K=\frac{ \int d^d x \sqrt{\pm g(x)} K(x,x;s)}{\int d^d x \sqrt{\pm g(x)}} \sim \frac{1}{(4 \pi s)^{d/2}} \sum_{n=0}^{\infty} a_n s^n \,.
\end{equation}

The solution to \eqref{hke} is formally provided by the expression $K(x,y;s)=\langle x | e^{-s \Delta} | y\rangle$. Equivalently, one can introduce a complete basis of eigenfunctions of the Laplacian operator $\Delta$, namely $\{ \phi_j(x)\}$, with eigenvalues $\lambda_j$ and such that 
\begin{equation}
K(x,y;s)=\sum_j e^{-\lambda_j s}  \phi_j(x)\phi^*_j(x)\,,
\end{equation}
which in flat space-time $g_{\mu\nu}\!=\!\eta_{\mu\nu}$ acquires the usual expression
\begin{equation}
K_\eta(x,y;s)=\int \frac{d^d p}{(2 \pi)^d}\, e^{-p^2 s}  \,  e^{\imath p\cdot (x-y)} \,.
\end{equation}
These latter relations, and in particular the formal expression $K(x,y;s)=\langle x | e^{-s \Delta} | y \rangle$, make also explicit the link to the propagator in quantum mechanics, and suggest the way it can be extended covariantly to relativistic systems. In Minkowski space-time, for relativistic continuous fields, the definition of the Feynman propagator $D_{F}(x,y)$ traces back to introduction of the so-called Schwinger proper-time
\begin{eqnarray}
D_F(x,y)&=& \int \frac{d^d p}{(2\pi)^d}  e^{-\imath p \dot (x-y)}\, \frac{\imath}{p^2-m^2 +\imath \epsilon} 
\nonumber\\
&=& \int \frac{d^d p}{(2\pi)^d}  e^{-\imath p \dot (x-y)}\, 
\int_0^\infty ds e^{\imath s (p^2-m^2 +\imath \epsilon)}
\nonumber\\
&=& \int_0^\infty \!\! ds \, e^{\imath s (p^2-m^2 +\imath \epsilon)} \int \frac{d^d p}{(2\pi)^d} e^{-\imath p \dot (x-y)}
\nonumber\\
&=& \int_0^\infty \!\! ds \, e^{- s \epsilon}\, e^{- \imath s m^2} K_\eta(x,y;-\imath s)\,.
\end{eqnarray}
This allows for a double useful identification, at least working on flat space-time. The Schwinger proper-time corresponds to the Wick-rotation of the thermal time, and hence the Feynman propagator to the integration of the heat-kernel. At the same time, the Schwinger proper-time is by definition conjugated to the Casimir of the space-time, or in other words, it accounts for the relativistic structure of the manifold, introducing a concept of proper time which depends on the space-time element.  \\

On the other hand, the Ricci flow is precisely the heat equation of the metric on a \emph{d-dimensional} manifold (either Riemannian or pseudo-Riemannian), namely
\begin{equation}
\frac{d g_{\mu \nu}}{d s }= -2 R_{\mu \nu}\,.
\end{equation}
One of the main result of our analysis 
stems in having recovered an intimate link between the Ricci flow and the stochastic quantization. This thermal time, interpolating among different scales in the RG flow, has therefore the interpretation of stochastic time. 

\subsection{Conformal transformations and the role of the scale factor} \label{conf}
\noindent 
In order to provide a physical intuition of the conformal transformations at the base of the Ricci flow, we inspect the features connected to this symmetries in GR. Specifically, we move from the Einstein-Hilbert action of gravity, which encodes two transverse propagative degrees of freedom of integer spin $\pm2$. A further longitudinal scalar massless excitation can be unfrozen thanks to the introduction of the cosmological constant term in the action. The cosmological constant term can be indeed considered as part of the energy momentum tensor of the theory of gravity coupled to matter. The inspection of the second order perturbation of the Einstein equations of motion provides ground to this intuitive statement. Nonetheless, there is a surprise we are going to comment about in the following.

\subsubsection{Conformal ghost in the geometric stochastic flow}
\noindent 
The background solution in presence of the cosmological constant $\Lambda$, and in absence of other forms of matter, is provided by either the de Sitter or anti-de Sitter solution, respectively for either $\Lambda >0$ or $\Lambda < 0$, both maximally symmetric solutions of the Einstein equations. Denoting these solutions as $\bar{g}_{\mu \nu}$, and with $h_{\mu \nu}$ the perturbation, one easily finds that for the Einstein-Hilbert action          
\begin{eqnarray}
(\sqrt{-g}R)^{(2)} &=& \sqrt{-\bar{g}} \Big[ 
\bar{R} \left( -\frac{1}{4} h_{\mu \nu} h^{\mu \nu} + \frac{1}{8} h^2 \right)  \nonumber   \\
&& + \bar{R}_{\mu \nu} \left( h^{\mu}_{\rho} h^{\nu \rho} - \frac{1}{2} h h^{\mu \nu} \right)  \nonumber  \\
&& + \frac{1}{4} \bar{\nabla}_\mu h\,  \bar{\nabla}^\mu h - \frac{1}{2} \bar{\nabla}_\mu h\,  \bar{\nabla}_\nu h^{\mu \nu} \nonumber  \\
&& - \frac{1}{4} \bar{\nabla}_\rho h^{\mu \nu}  \bar{\nabla}^\rho h_{\mu \nu} + \frac{1}{2}  \bar{\nabla}_\rho h_{\mu \nu}  \bar{\nabla}^\rho h^{\mu \nu} \Big] \nonumber  \\
&& + {\rm total \ \ derivative}\,.  \nonumber 
\end{eqnarray}

Decomposing the metric --- see e.g.  \cite{Alvarez-Gaume:2015rwa} --- in its irreducible components
\begin{eqnarray}
h_{\mu \nu}&=& h_{\mu \nu}^\perp + \partial_\mu a^\perp_\nu +  \partial_\nu a^\perp_\mu \\
&&+ \left( \partial_\mu \partial_\nu  -\frac{1}{4} \eta_{\mu \nu}  \Box \right) a + \frac{1}{4} \eta_{\mu \nu} \varphi\,, \nonumber 
\end{eqnarray}
with $h^\perp$ transverse traceless
\begin{eqnarray}
\partial^\mu h_{\mu \nu}^\perp = \eta^{\mu \nu} h_{\mu \nu}^\perp =0\,,
\end{eqnarray}
$a_{\mu}^\perp$ transverse, thus 
\begin{eqnarray}
\partial^\mu a_{\mu }^\perp =0\,,
\end{eqnarray}
$a$ scalar and $\varphi=h^\mu_\mu$ trace of the perturbation tensor, one obtains 
\begin{eqnarray} \label{pehf}
\mathcal{S}_{\rm EH}^{(2)}&=& \frac{c^3}{16 \pi G} \int d^4x \sqrt{-\bar{g}}  \\
&& \times \left[ \frac{1}{4} h_\perp^{\mu \nu} \left(\bar{\Box} - \frac{\bar{R}}{6} \right) h^\perp_{\mu \nu} -\frac{3}{32} \Phi \left(\bar{\Box} + \frac{\bar{R}}{3} \right) \Phi
\right] \,, \nonumber 
\end{eqnarray}
in which $\Phi=\varphi-\bar{\Box}a$ represents a scalar massless excitation, invariant under the gauge transformations
\begin{eqnarray}\label{git}
&& h^\perp_{\mu \nu} \rightarrow h^\perp_{\mu \nu}\,, \\
&& a^\perp_\mu \rightarrow a^\perp_\mu  + \xi^\perp_\mu \,,\nonumber \\
&& a \rightarrow a + 2 \xi \,,\nonumber \\
&& \varphi \rightarrow \varphi + 2 \box \xi\,,  \nonumber 
\end{eqnarray}
namely $\Phi=\varphi- \Box a \rightarrow \Phi$ under Eq.~\eqref{git}.\\

Despite the appearance of a quadratic kinetic term in Eq.~\eqref{pehf}, this the excitation connected to $\Phi$ is not a physical degree of freedom but rather a ghost, as evident from the opposite sign with respect to the kinetic term of the graviton.

Nonetheless, this ghost is harmless, since is not a propagative degree of freedom. This can be understood from its non-locality, namely  
\begin{eqnarray}
\Phi=P^{\mu \nu} h_{\mu \nu}\,,  
\end{eqnarray}
with 
\begin{eqnarray}
P^{\mu \nu}= \eta^{\mu \nu} - \frac{4}{3} \Box^{-1} \left( \partial^\mu \partial^\nu -\frac{1}{4} \eta^{\mu \nu} \Box \right) \,.
\end{eqnarray}
It immediately follows indeed that the Cauchy problem cannot be solved within this framework, since the initial data at a given time for the metric perturbations $h_{\mu \nu}$ are not enough to determine $\Phi$. This confirms the non existence of an extra propagative degree of freedom. \\

An alternative way to unveil the real nature of the scalar ghost excitation consists in performing a residual gauge transformation on the gauge invariant quantities $h_{\mu \nu}^\perp$ and $\Phi$, which reads
\begin{eqnarray}
h_{\mu \nu}^\perp \rightarrow h_{\mu \nu}^\perp + \nabla_\mu \xi_\nu +  \nabla_\nu \xi_\mu \,, \quad \Phi \rightarrow \Phi + 2 \nabla^\mu k_\mu \,, 
\end{eqnarray}
with $\xi_\mu$ and $k_\mu$ conformal Killing vector that fulfil  
\begin{eqnarray}
&& \nabla^\mu \xi_\mu =0\,, \quad \left( \Box + \frac{R}{4}\right) \xi_\mu =0 \,, \\
&& \nabla_\mu k_\nu + 4 \nabla_\nu k_\mu = \frac{1}{2} g_{\mu \nu} \nabla^\sigma k_\sigma \,. \nonumber 
\end{eqnarray}
Then $\xi_\mu$ eliminates four of the six components of $h^\perp_{\mu \nu}$, leaving only two propagating components corresponding to the helicity states of the graviton, while the divergence of the conformal Killing vector eliminates the spin $0$ ghost scalar excitation $\Phi$.\\

It is now crucial to notice that a thermal flow of the metric driven either by the Ricci tensor, or by the difference of the Ricci tensor and the target matter tensor, breaks the aforementioned residual gauge symmetry. Asymptotically, at equilibrium, when the Einstein equations hold, the residual gauge symmetry is recovered, and the ghost mode can be finally reabsorbed. Nonetheless, outside of the equilibrium, the scalar ghost mode arises in the geometric flow as a non-local propagative degree of freedom. We finally notice that in a stochastic geometric flow, endowed {\it ad hoc} with  of any non-vanishing space-time curvature term, a similar effect to the one we have just shown is found.

\subsection{Finite theories of quantum gravity and RG-flow}
\label{Stelle}
\noindent 
The fascinating possibility to recover a dynamical breakdown of the conformal invariance suggests to make contact with conformal invariant theories of gravity, among which prominently appears Stelle's theory \cite{Stelle:1977ry}. Considering this latter may indeed provide particular advantage for the RG flow program we are implementing here, based on the Ricci flow equations. In our stochastic quantization framework the appearance of ghosts, which plagues the Stelle theory --- this would be otherwise a finite theory of quantum gravity --- might be indeed removed due to the independence on the gauge-fixing.\\ 

This picture provides an unprecedented pathway toward the quantization of gravity. Besides the possibility to obtain a finite and unitary theory of quantum gravity, solving a longstanding problem in physics, a further most appealing perspective appears. The emergence in the UV regime of a conformal symmetry hinges towards the natural convergence to zero of the theory's coupling constants, including the gravitational one. This drives the theory toward a topological phase. Moving from this latter towards the infrared regime, due to the dynamical conformal symmetry breaking, theories accounting for propagating degrees of freedom can be finally recovered. \\

This scenario is somehow reminiscent of the dimensional transmutation in QCD, in which the $\Lambda_{\rm QCD}$ scale is recovered as a signature of the conformal anomaly. Nonetheless, the higher order derivative terms that are present in the Stelle theory naturally encode several additional dynamical degrees of freedom. Among these latter, in the \cite{Stelle:1977ry}, we can identify scalar modes that can be associated to the Higgs iso-doublet field. The conformal symmetry breaking mechanism can be then conjectured to be instantiated by degrees of freedom that can be then identified with those ones of the Higgs field. At the same time, within this framework the very same emergence of the arrow of time can be traced back to the conformal symmetry breaking. \\

\bibliography{refs}
\end{document}